\documentclass[letterpaper, 10 pt, conference]{ieeeconf}  % Comment this line out
\IEEEoverridecommandlockouts                              % This command is only
\usepackage{graphicx} % for pdf, bitmapped graphics files
\usepackage{amsmath} % assumes amsmath package installed
\usepackage{amssymb}  % assumes amsmath package installed
\usepackage{subfig}
\usepackage{cite}

\makeatletter
\def\Hline{%
\noalign{\ifnum0=`}\fi\hrule \@height 2pt \futurelet
\reserved@a\@xhline}
\makeatother

\def\figref#1{Fig.~\ref{#1}}

\def\figsssref#1#2#3#4{Figs.~\ref{#1}-\ref{#4}}
\def\tabref#1{Table~\ref{#1}}
\def\secref#1{Section~\ref{#1}}
\def\eqref#1{eq.~(\ref{#1})}

\begin{document}

\title{\LARGE \bf
Deep demosaicking for multispectral filter arrays
}

\author{Kazuma Shinoda$^{1}$, Shoichiro Yoshiba$^{1}$ and Madoka Hasegawa$^{1}$% <-this % stops a space
\thanks{$^{1}$The authors are with Graduate School of Engineering, Utsunomiya University, 7-1-2 Yoto, Utsunomiya, Japan}%
}

\maketitle
\thispagestyle{empty}
\pagestyle{empty}

%%%%%%%%%%%%%%%%%%%%%%%%%%%%%%%%%%%%%%%%%%%%%%%%%%%%%%%%%%%%%%%%%%%%%%%%%%%%%%%%
\begin{abstract}
We propose a novel demosaicking method for multispectral filter arrays based on a deep convolutional neural network. The proposed method first interpolates mosaicked multispectral images utilizing a bilinear approach, then applies a residual network to initial demosaicked images. The residual network consists of various three-dimensional convolutional layers and a rectified linear unit for describing the features of a multispectral data cube. Experimental results reveal that the proposed method outperforms conventional demosaicking methods.
\end{abstract}

%%%%%%%%%%%%%%%%%%%%%%%%%%%%%%%%%%%%%%%%%%%%%%%%%%%%%%%%%%%%%%%%%%%%%%%%%%%%%%%%
\section{Introduction}

Multispectral images (MSIs) have been utilized in many remote sensing and medical applications \cite{JMBioucasDias2013,GLu2014}. A multispectral filter array (MSFA) is one solution for capturing an MSI in a single shot at low cost. The architecture of an MSFA has a different spectral filter for each pixel of an image. Images captured by an MSFA have only one value for each pixel, but a full-resolution MSI can be obtained by recovering the missing spectrum information. This recovering process is referred to as {\it demosaicking}. 

Various filter array patterns and demosaicking methods \cite{JBrauers2006,FYasuma2010,HKAggarwal2014,JMizutani2014,KShinoda2015,JJia2016,SOgawa2016,KShinoda2016,YMonno2015,SMihoubi2017,KShinoda2017,KShinoda2018} have been proposed for improving demosaicked image quality. Brauers et al. \cite{JBrauers2006} proposed a six-band MSFA arranged in 3 $\times$ 2 pixels in a straightforward manner for fast linear interpolation. Their demosaicking method first applies bilinear interpolation to each band. Inter-band correlations are then corrected by smoothing inter-band differences. Yasuma et al. \cite{FYasuma2010} designed a seven-band MSFA composed of three primary color filters and four secondary color filters. A simple four-band MSFA was proposed by Aggarwal et al. \cite{HKAggarwal2014}, who presented a versatile demosaicking method based on an $l_{1}$-norm minimization problem. Jia et al. \cite{JJia2016} proposed a Fourier spectral filter array utilizing a Fabry-Perot-type filer to remove aliasing. Monno et al. \cite{YMonno2015} proposed a five-band MSFA and determined that the sampling density of G-band data was higher than that of the other spectral bands because the human eye is more sensitive to the G-band compared to other spectral bands. Mihoubi et al. \cite{SMihoubi2017} proposed a demosaicking method based on the correlation between each spectral band and a pseudo-panchromatic image (PPI), which represents the average image over all bands. In our previous work, we designed an MSFA pattern for medical applications based on training data \cite{KShinoda2015,KShinoda2018}. Because the number of non-measured pixels in an MSFA is larger than that in an RGB Bayer color filter array, a more accurate demosaicking method is required.

An RGB demosaicking method based on a convolutional neural network (CNN) was proposed in \cite{DSTan2018} and achieved excellent performance. This method first produces a rough demosaicked image by applying a conventional RGB demosaicking method (Gradient-Corrected Bilinear Interpolation, GCBI\cite{HSMalvar2004}),
then reduces color artifacts by utilizing deep residual estimation and multi-model fusion. 

In this paper, we propose a novel demosaicking method for MSIs based on a deep CNN. The proposed method first interpolates a mosaicked image utilizing a bilinear approach, then reduces artifacts in the tentative demosaicked image utilizing a deep residual network (ResNet) \cite{KHe2016} and three-dimensional (3D) convolutions \cite{YLi2017}. It is expected that the ResNet can reduce the error between the initial demosaicked image and training data, and the 3D-CNN can learn local signal changes in both the spatial and spectral dimensions of feature cubes effectively. We compared the performance of our demosaicking method to that of a bilinear approach and PPI difference (PPID) method \cite{SMihoubi2017}.

This remainder of paper is organized as follows. In \secref{sec:level2}, we describe our proposed demosaicking method. We present experiments to test our method and their results in \secref{sec:level3}. \secref{sec:level4} contains our conclusions.
%%%%%%%%%%%%%%%%%%%%%%%%%%%%%%%%%%%%%%%%%%%%%%%%%%%%%%%%%%%%%%%%%%%%%%%%%%%%%%%%
\section{Proposed demosaicking method} \label{sec:level2}

\begin{figure}[tb]
 \centering
  \includegraphics[width=1\linewidth]{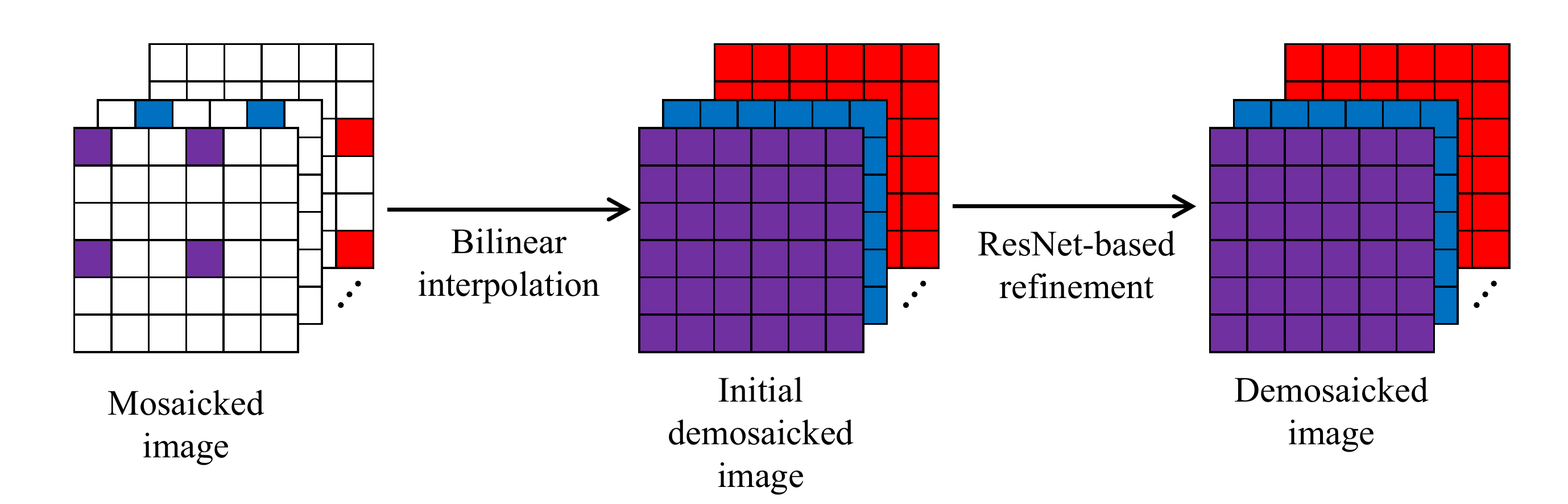}
 \caption{Proposed multispectral demosaicking flow.}
 \label{flow}
\end{figure}

\begin{figure}[tb]
 \centering
  \includegraphics[width=0.8\linewidth]{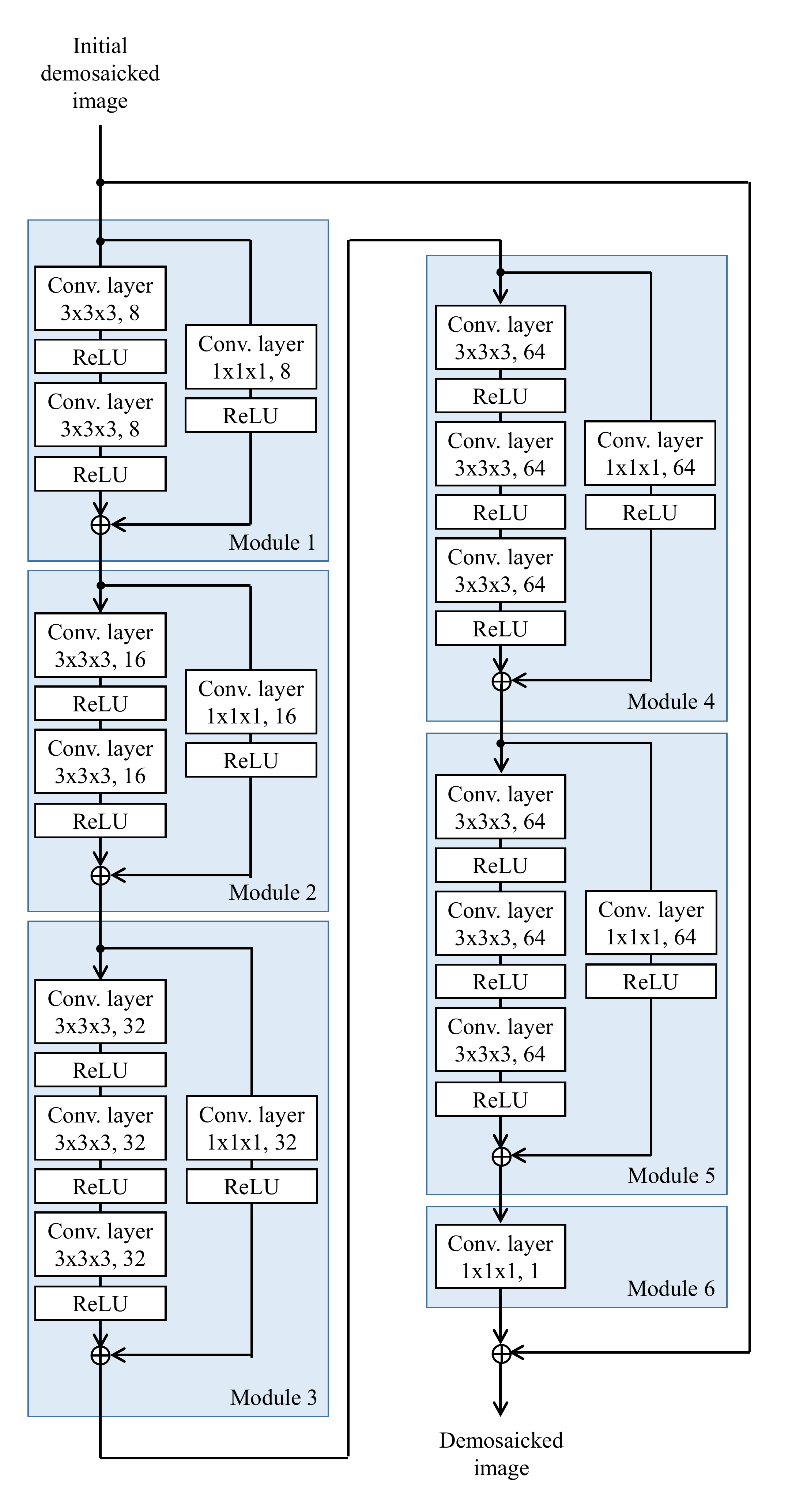}
 \caption{Network structure of the proposed method.}
 \label{network}
\end{figure}

\begin{figure}[tb]
 \centering
  \includegraphics[width=0.85\linewidth]{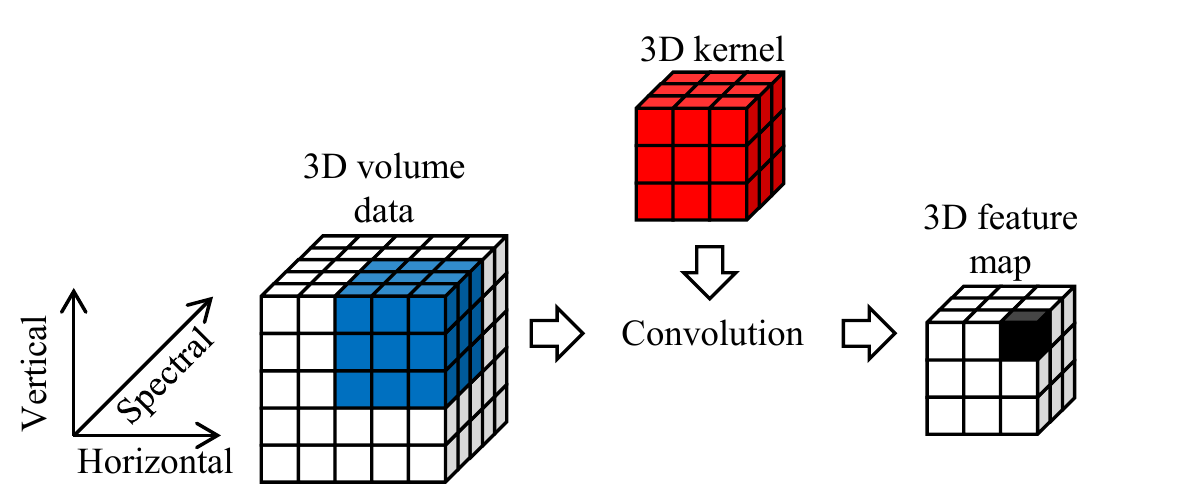}
 \caption{3D convolution in the proposed network.}
 \label{3DCNN}
\end{figure}

The proposed method consists of two steps: initial demosaicking and ResNet-based refinement, as shown in \figref{flow}. The first step interpolates a mosaicked image utilizing a bilinear approach with an $N \times N$ window ($N$ is the block size of MSFA plus one). This bilinear interpolation is performed for each band independently. Next, the second step refines the initial demosaicked image by utilizing ResNets and 3D-CNN. The proposed network structure is illustrated in \figref{network}. Here, $\oplus$ is the addition operator for each pixel and ($x \times y \times z, n)$ in the convolutional layer is the 3D filter size $(x, y, z)$ in horizontal-vertical-spectral directions, where the number of filter types is $n$. The proposed network consists of six modules, where each module from one to five has a convolutional layer for the 3D-CNN with a rectified linear unit (ReLU) and shortcut connection. Note that the shortcut connection for each module includes a $1 \times 1 \times 1$ convolutional layer because the number of output features for each module is different from the number of input features. The final convolutional layer in module six combines 32 features into a single data cube (i.e., produces a residual). Next, the final demosaicked image is obtained by adding the initial demosaicked image to the residual data. The longest shortcut connection (from the beginning to the end of the network) does not include a convolutional layer.

The procedure for 3D convolution is presented in \figref{3DCNN}. Note that the obtained feature map from the proposed convolutional layer is a 3D data cube, rather than a 2D structure. The number of spectral bands in the kernel does not equal that in the input data cube and the convolution process is performed along both spatial and spectral directions. This 3D convolution can extract local signal changes in both the spatial and spectral dimensions of feature cubes effectively in the case of MSIs \cite{YLi2017}. In the convolutional layer of the proposed network, a $1 \times 1 \times 1$ or $3 \times 3 \times 3$ kernel is used for convolution. All kernel coefficients are determined by learning. The ReLU function in the proposed network is $f = {\rm max}(x, 0)$, which is expected to speed up training \cite{AKrizhevsky2012}. 

Each module in the proposed network is based on a ResNet structure. Although ResNets have been applied to comparatively deeper networks for solving the degradation problem \cite{KHe2016} by shortcutting input data, wider (rather than deeper) networks have shown similar or better results than those obtained from very deep networks \cite{SZagoruyko2016}. The proposed network is comparatively shallow, but the shortcut connection may have an effect on solving the degradation problem. The longest shortcut connection has the potential to suppress artifacts in demosaicked images effectively \cite{DSTan2018}.

%%%%%%%%%%%%%%%%%%%%%%%%%%%%%%%%%%%%%%%%%%%%%%%%%%%%%%%%%%%%%%%%%%%%%%%%%%%%%%%%
\section{Experimental results} \label{sec:level3}

\begin{figure}[tb]
 \centering
  \includegraphics[width=0.35\linewidth]{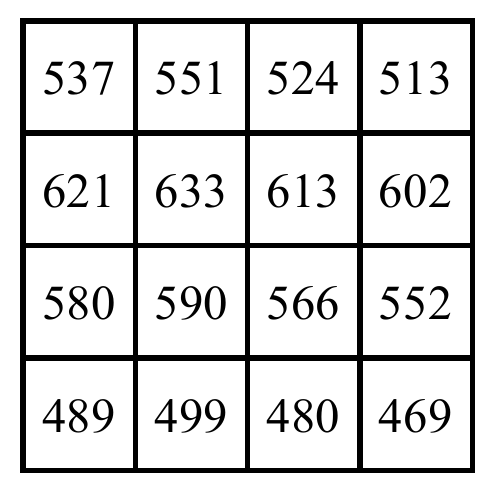}
 \caption{Multispectral filter array pattern. Each number indicates an optical center wavelength (nm).}
 \label{fig:MSFA}
\end{figure}

In our experiment, we used Python 2.7.13 and Keras 2.0.8 \cite{FChollet2015} (backend: Tensorflow 1.3.0 \cite{MAbadi2015} with the Adam optimizer \cite{DPKingma2015}) for the proposed method and the CAVE database \cite{FYasuma2010} for test images. The illuminant of the test images is CIE D65. According to the mosaicking method in \cite{SMihoubi2017}, the mosaicked images were obtained by using a 16-band MSFA, as shown in \figref{fig:MSFA}. We compared the peak signal-to-noise ratios (PSNRs) and appearances of the demosaicked images to those from the bilinear interpolation and PPID \cite{SMihoubi2017}.

In the proposed method, 8-fold cross validation was used for evaluating PSNR. This validation first divided 32 test images into eight groups evenly, of which seven groups were used for ResNet training 300 times and the remaining group was used for evaluation. Additionally, we divided each image into 16 sub-images (batch size of 8) for training to save memory consumption and computational time.

\begin{table}[t]
\centering
\caption{PSNR (dB) of demosaicked images}
 \label{tab:PSNR}
  \begin{tabular}{cccc} \hline
 & Bilinear & PPID & Proposed \\ \hline
 fake and real peppers & 35.77 & 43.02 & {\bf 47.43} \\ 
clay & 37.39 & 41.29 & {\bf 48.38} \\ 
beads & 26.81 & 29.70 & {\bf 32.69} \\
fake and real tomatoes & 33.81 & {\bf 42.76} & 41.52 \\
Average of 32 images & 34.58 & 40.38 & {\bf 43.05} \\ \hline
 \end{tabular}
\end{table}

The PSNRs of the demosaicked images are listed in \tabref{tab:PSNR}. The average PSNR of the proposed method was higher than that of both the bilinear interpolation and PPID methods, with a gain of 2.67 dB. This result indicates that the proposed ResNet-based refinement method can improve the quality of the initial images demosaicked via bilinear interpolation, and its PSNR exceeds that of conventional multispectral demosaicking methods, such as PPID. The number of images for which PPID achieved the highest PSNR was only two (including ``fake and real tomatoes demosaicked'') out of 32.

\begin{figure*}[tb]
 \centerline{
  \subfloat[Original]{\includegraphics[width=0.2\linewidth]{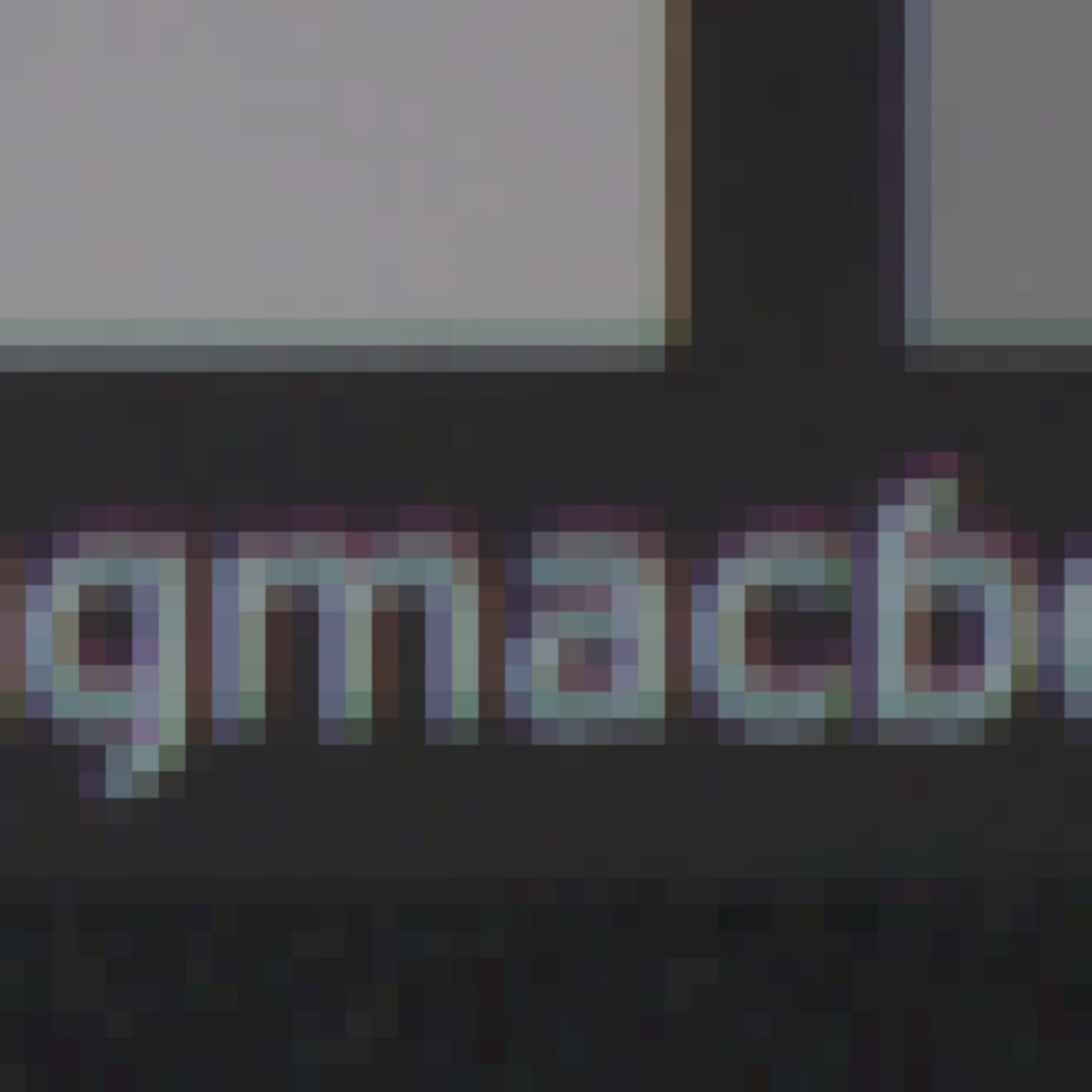}%
  \label{fig:fake_and_real_peppers_original}}
  \hfil
  \subfloat[Bilinear]{\includegraphics[width=0.2\linewidth]{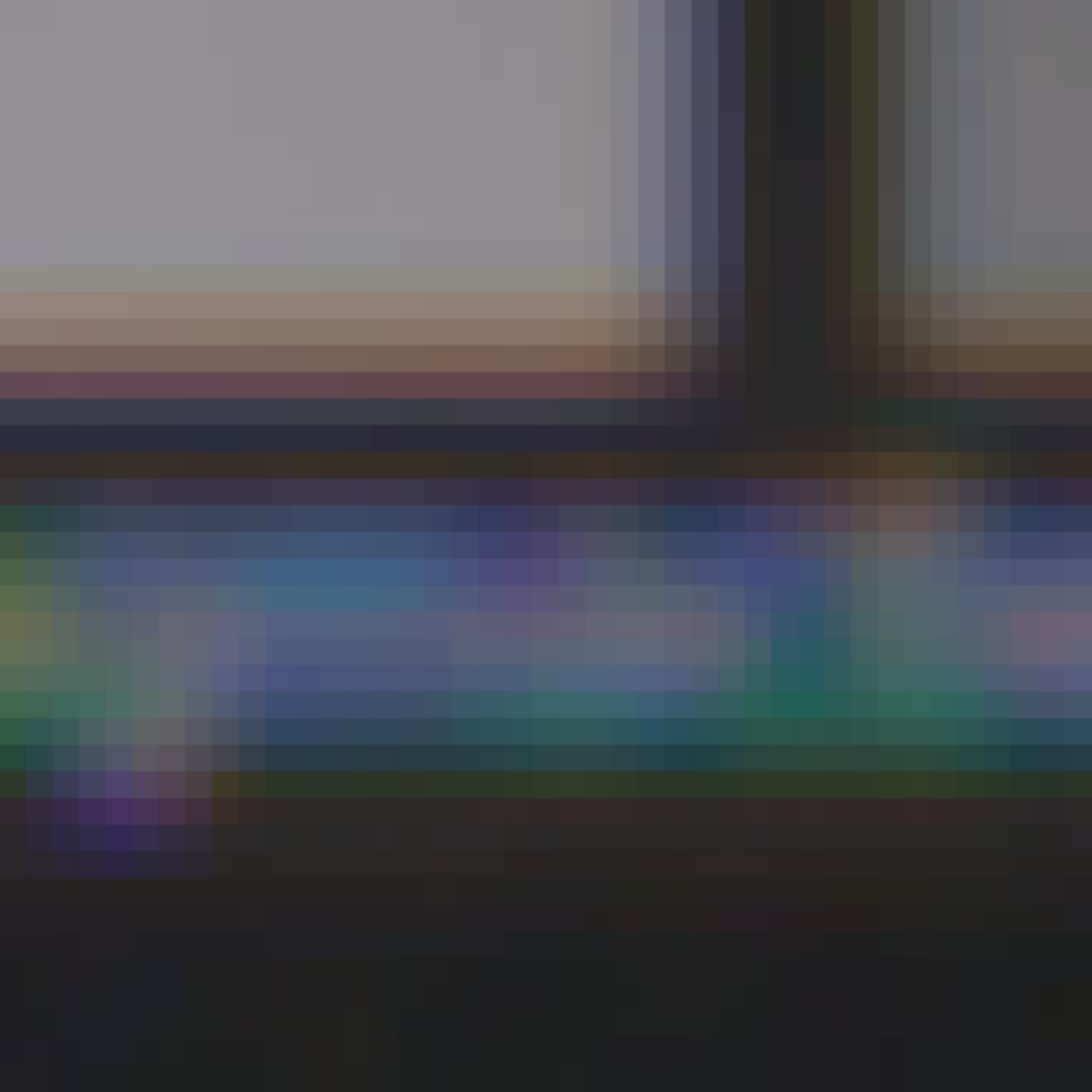}%
  \label{fig:fake_and_real_peppers_bilinear}}
  \hfil
  \subfloat[PPID]{\includegraphics[width=0.2\linewidth]{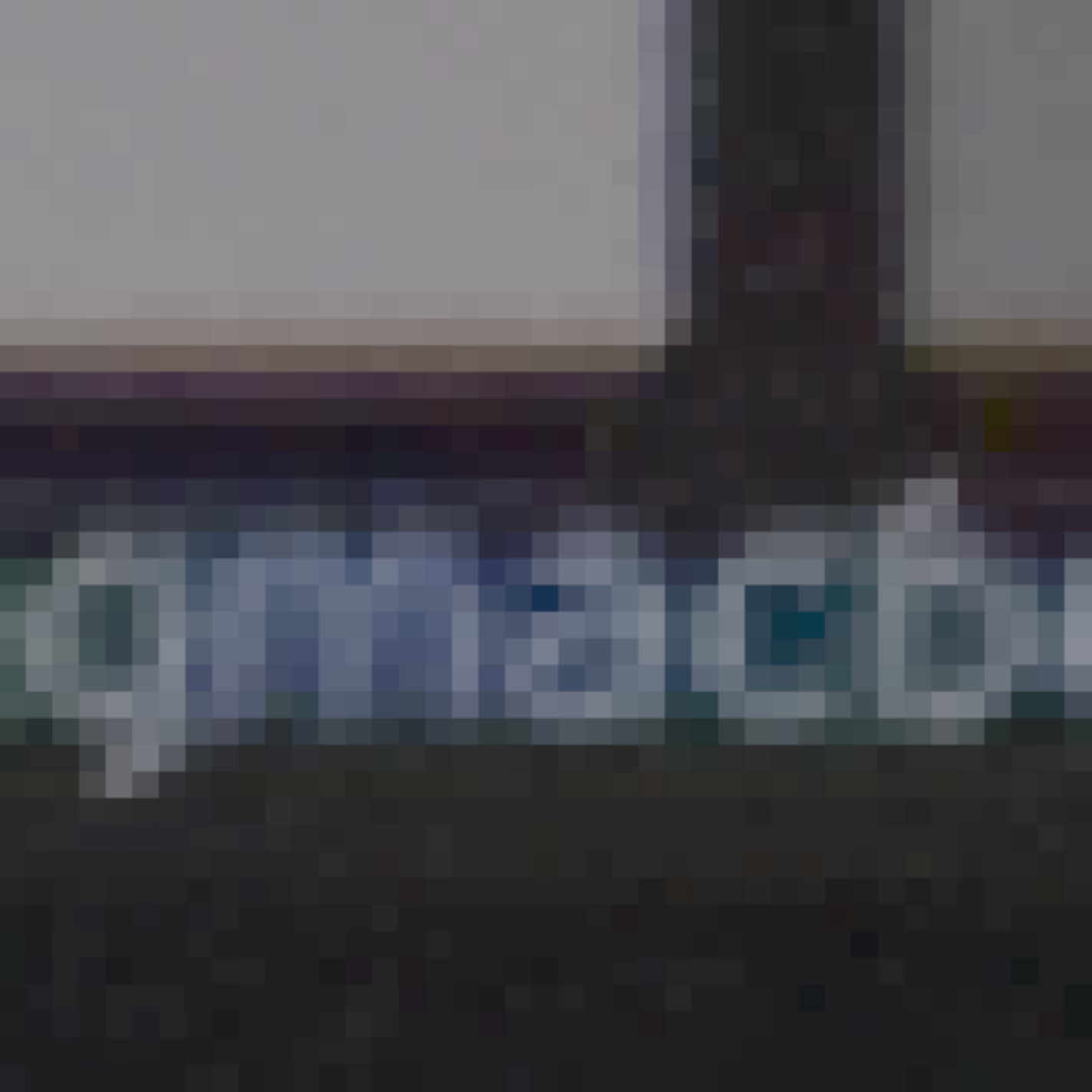}%
  \label{fig:fake_and_real_peppers_PPID}}
  \hfil
  \subfloat[Proposed]{\includegraphics[width=0.2\linewidth]{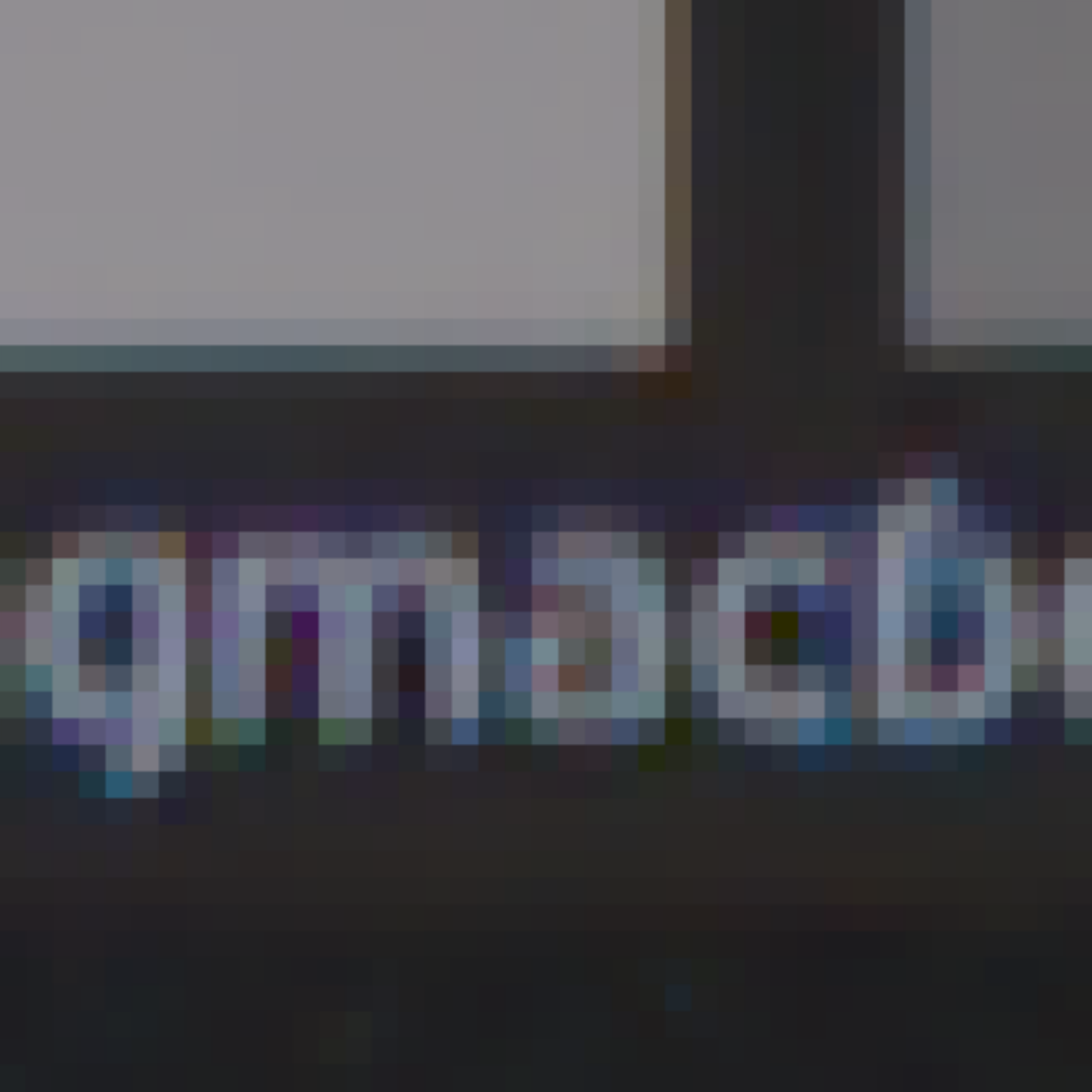}%
  \label{fig:fake_and_real_peppers_proposed}}
  }  
  \caption{Demosaicked images of ``fake and real peppers.''}
 \label{fig:fake_and_real_peppers_demosaicked}
\end{figure*}

\begin{figure*}[tb]
 \centerline{
  \subfloat[Original]{\includegraphics[width=0.2\linewidth]{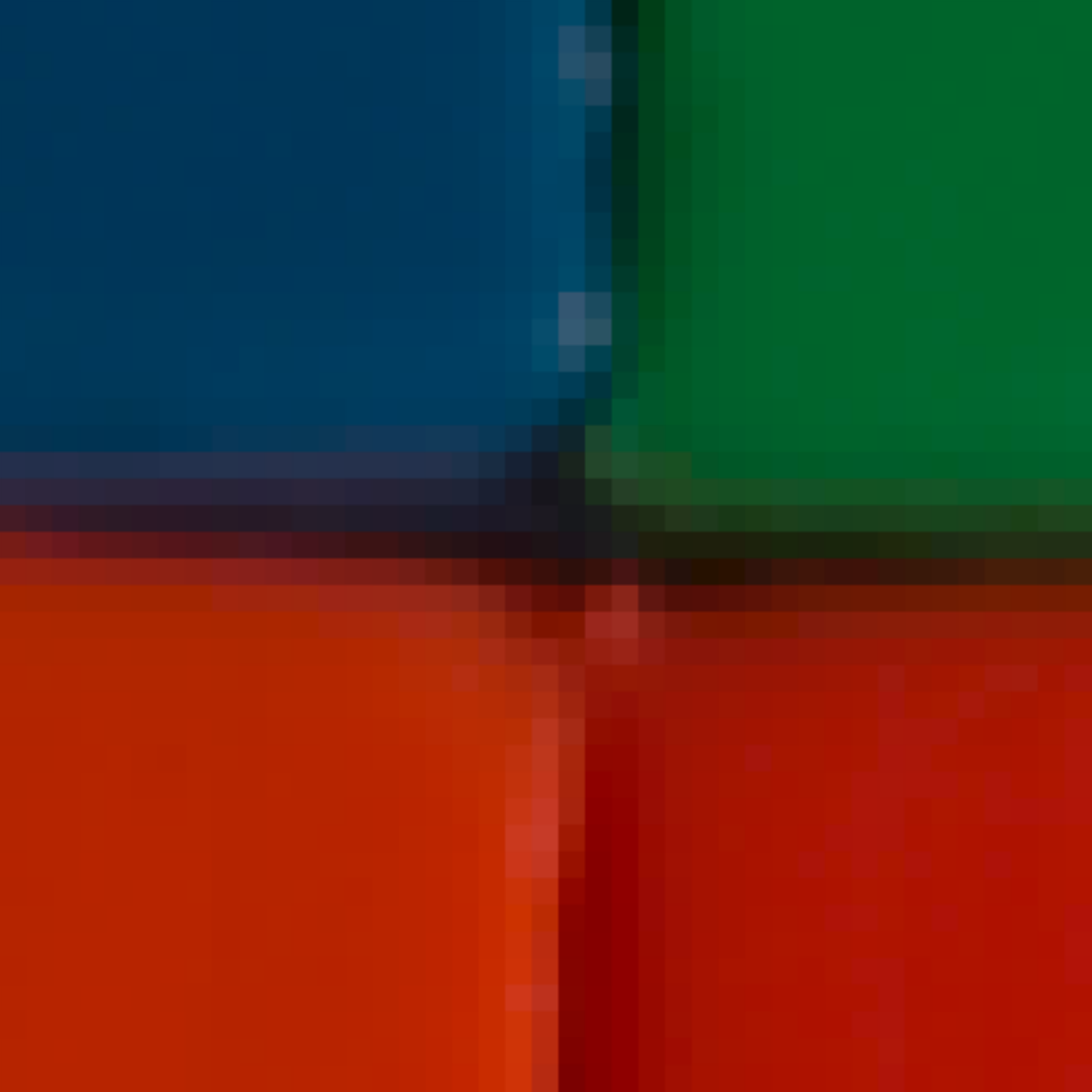}%
  \label{fig:clay_original}}
  \hfil
  \subfloat[Bilinear]{\includegraphics[width=0.2\linewidth]{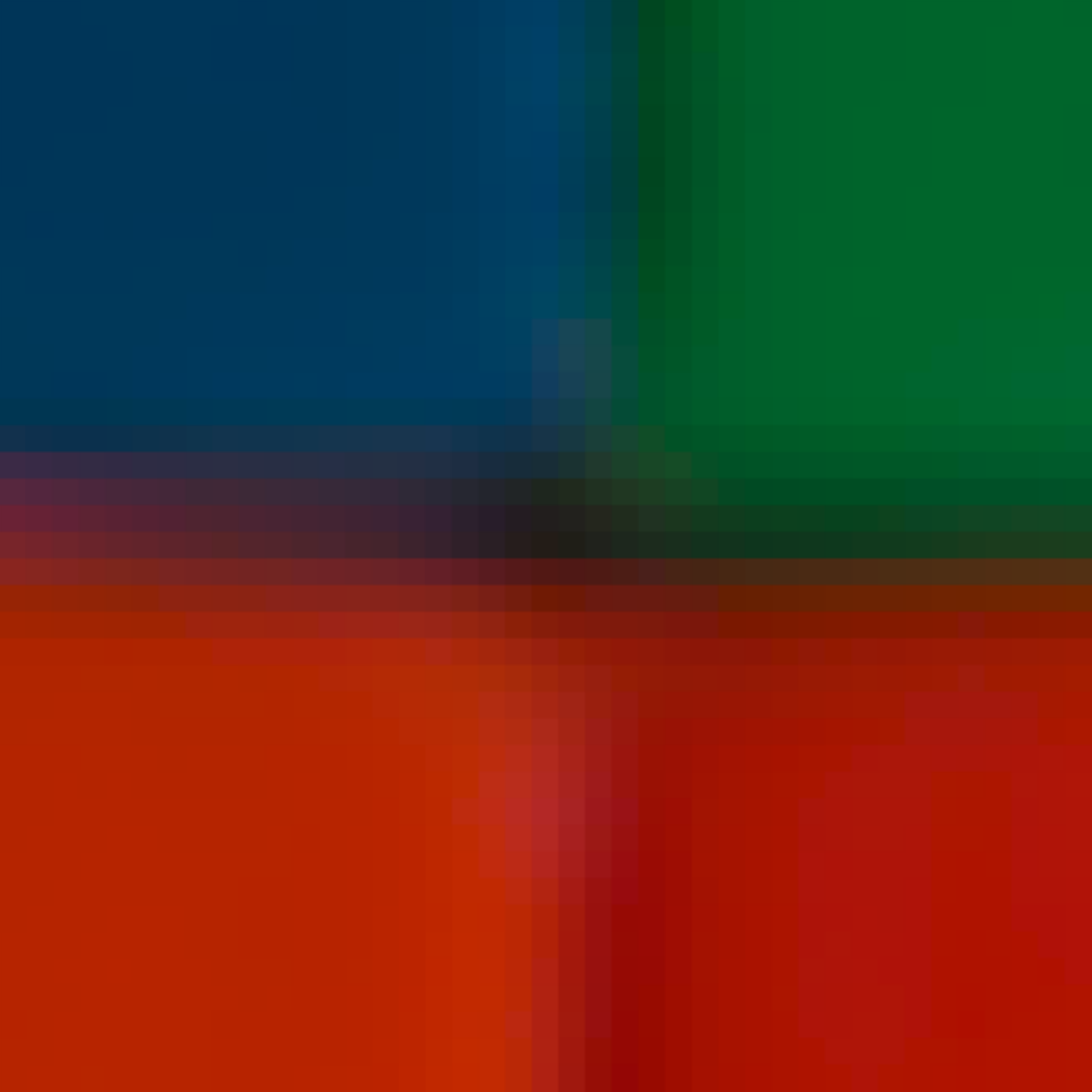}%
  \label{fig:clay_bilinear}}
  \hfil
  \subfloat[PPID]{\includegraphics[width=0.2\linewidth]{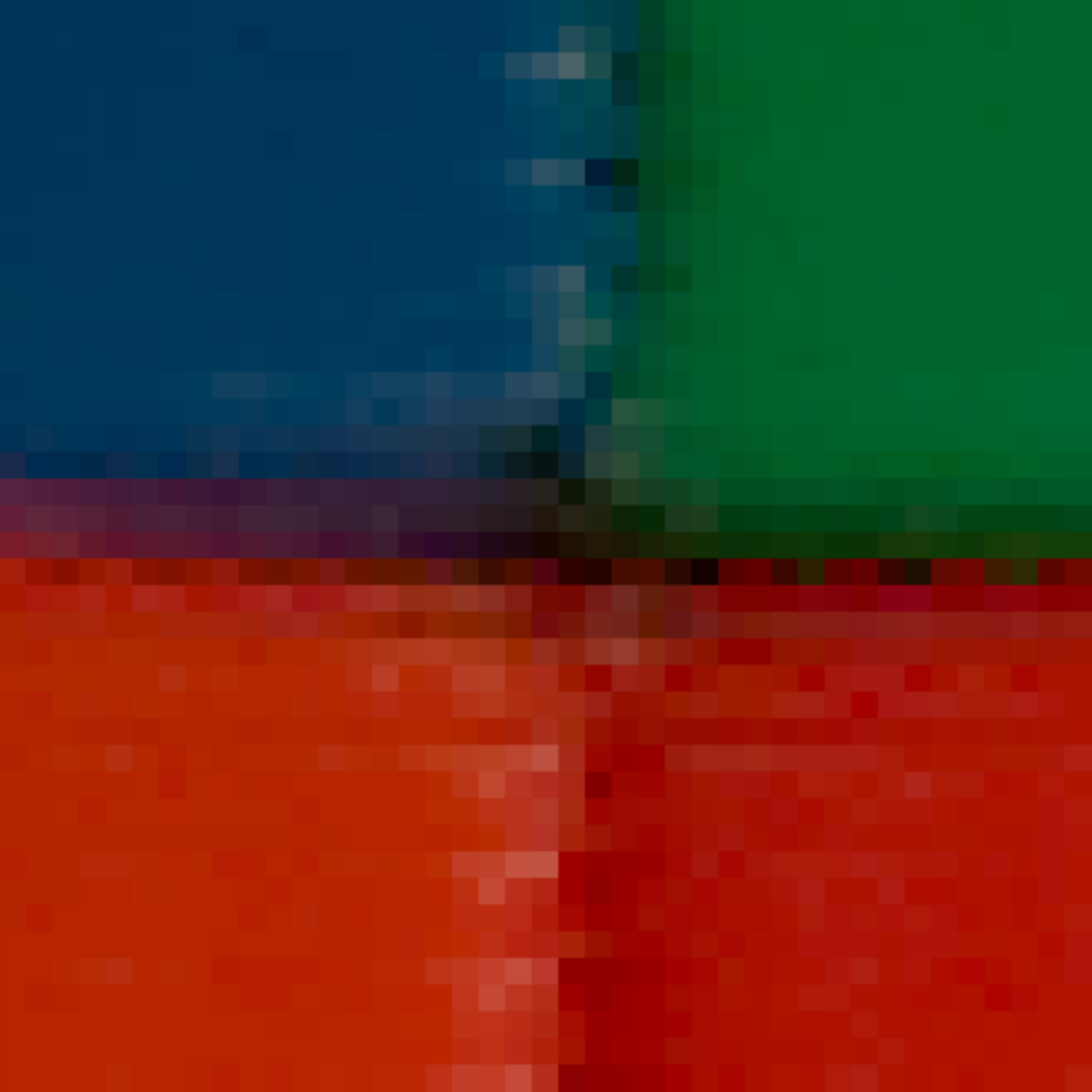}%
  \label{fig:clay_PPID}}
  \hfil
  \subfloat[Proposed]{\includegraphics[width=0.2\linewidth]{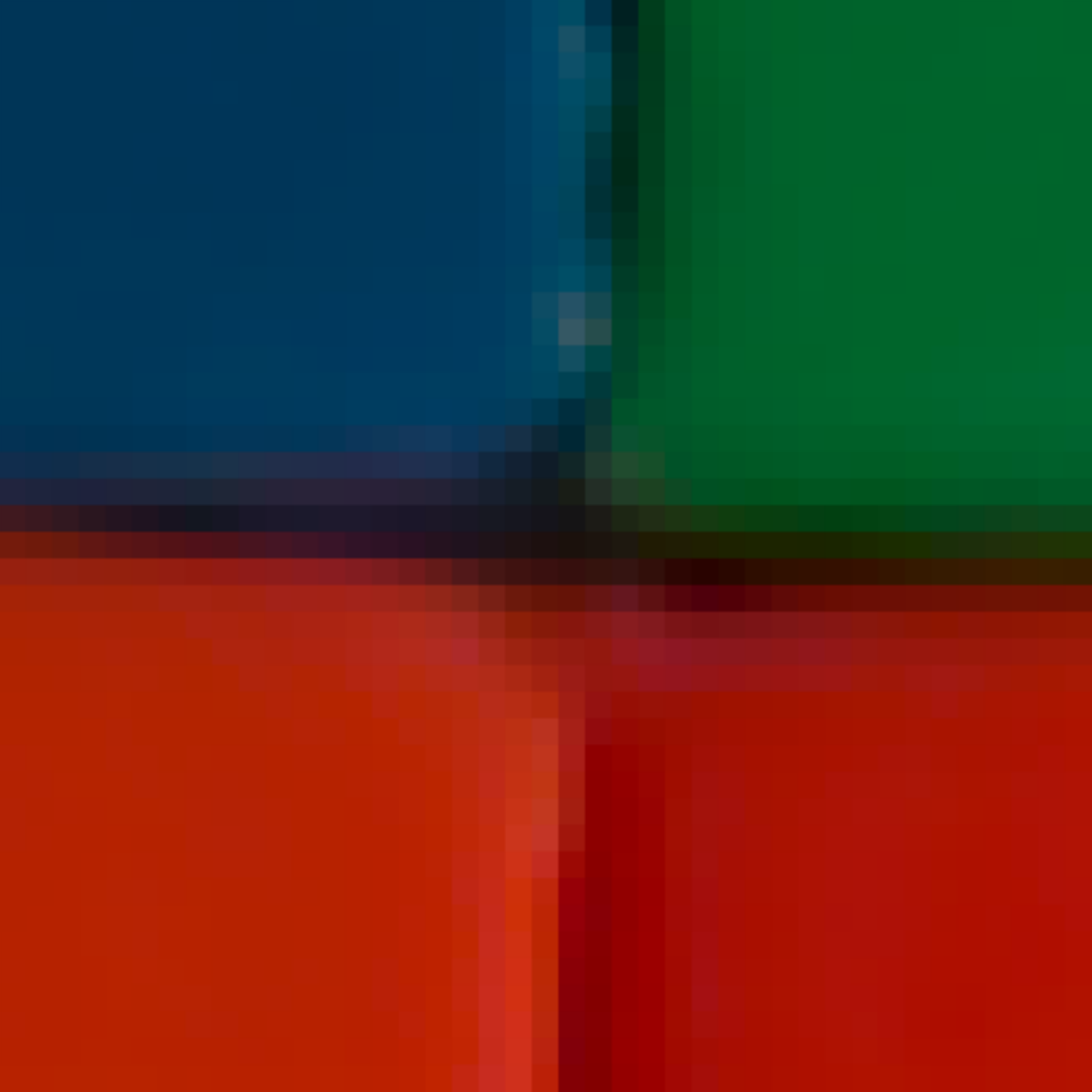}%
  \label{fig:clay_proposed}}
  }  
  \caption{Demosaicked images of ``clay.''}
 \label{fig:clay_demosaicked}
\end{figure*}

\begin{figure*}[tb]
 \centerline{
  \subfloat[Original]{\includegraphics[width=0.2\linewidth]{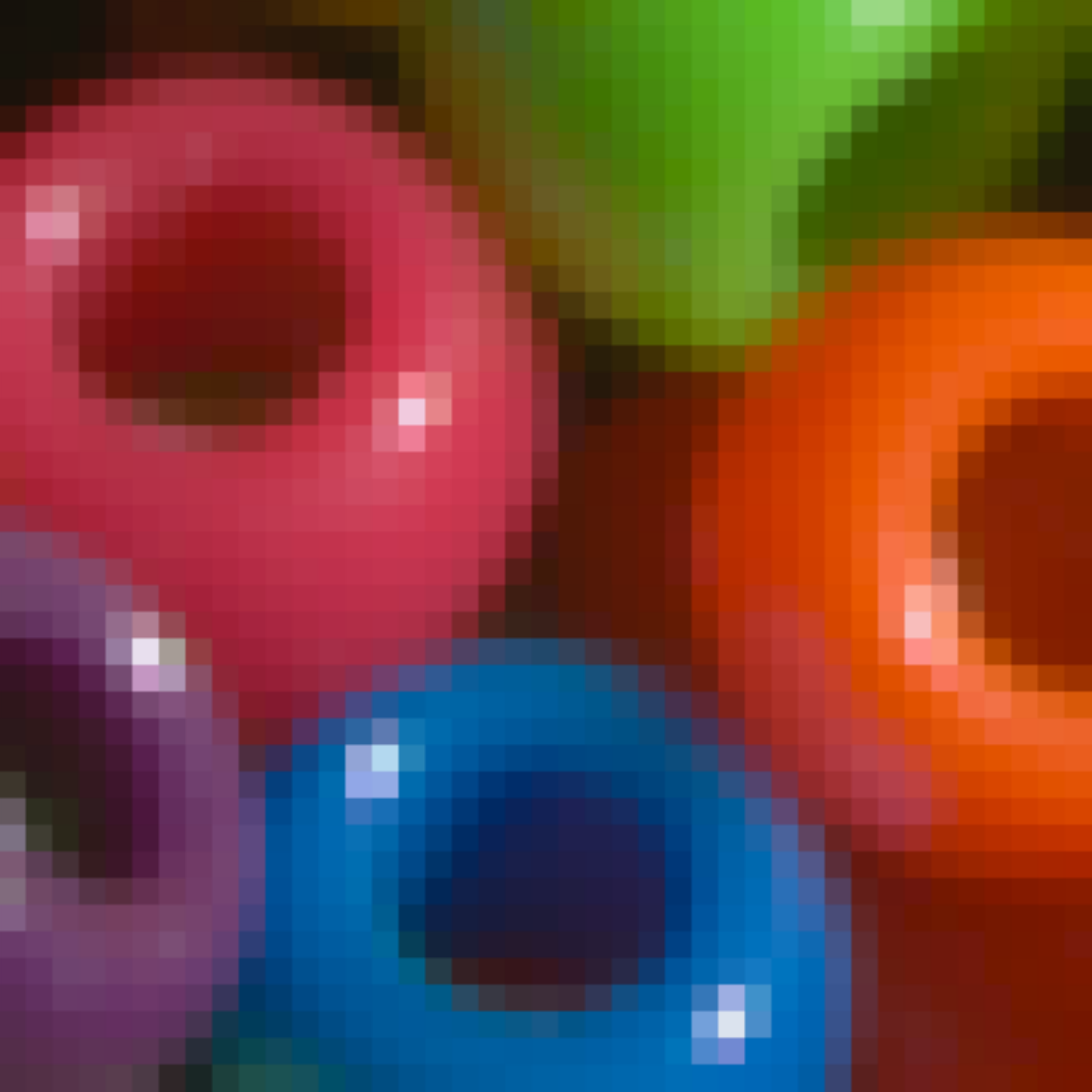}%
  \label{fig:beads_original}}
  \hfil
  \subfloat[Bilinear]{\includegraphics[width=0.2\linewidth]{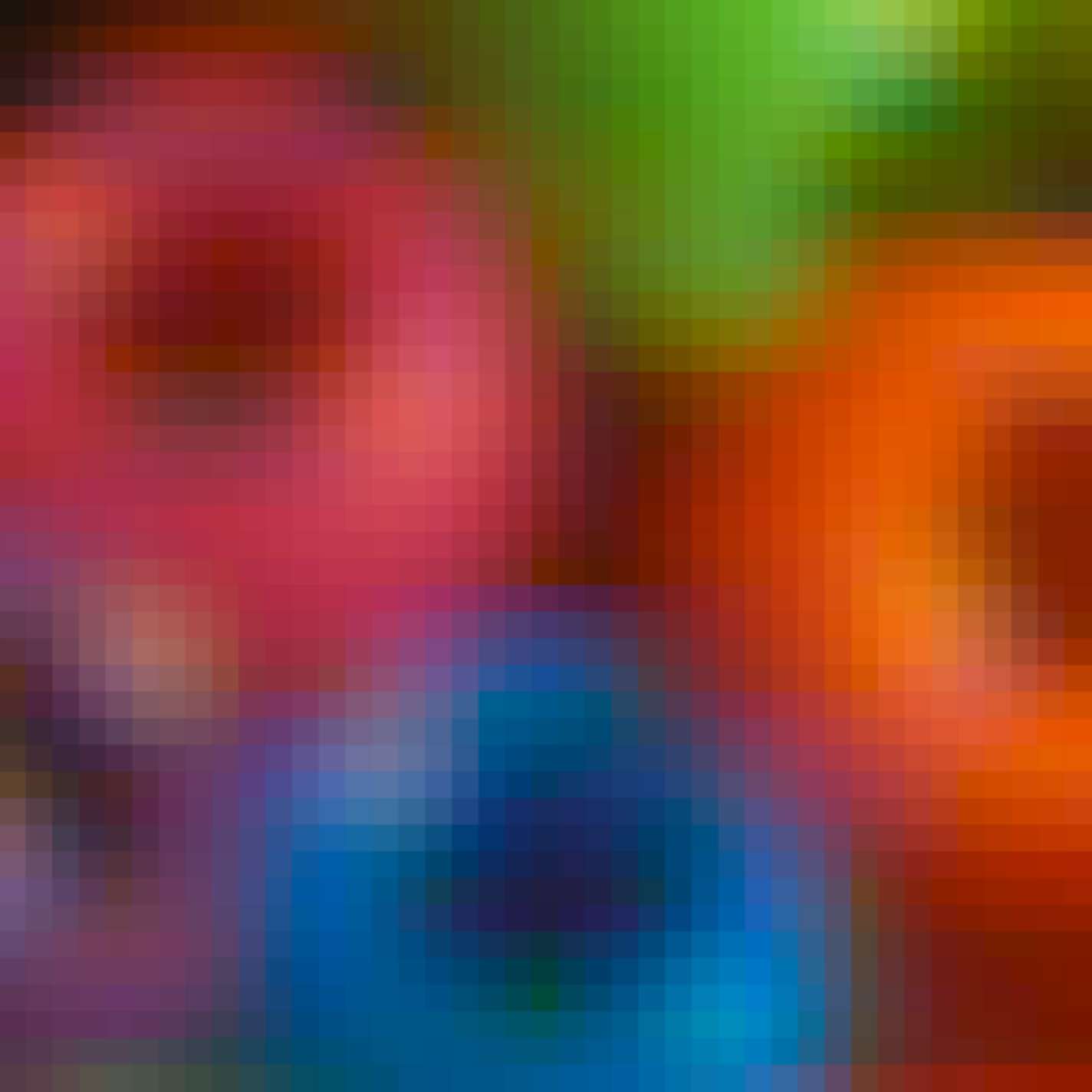}%
  \label{fig:beads_bilinear}}
  \hfil
  \subfloat[PPID]{\includegraphics[width=0.2\linewidth]{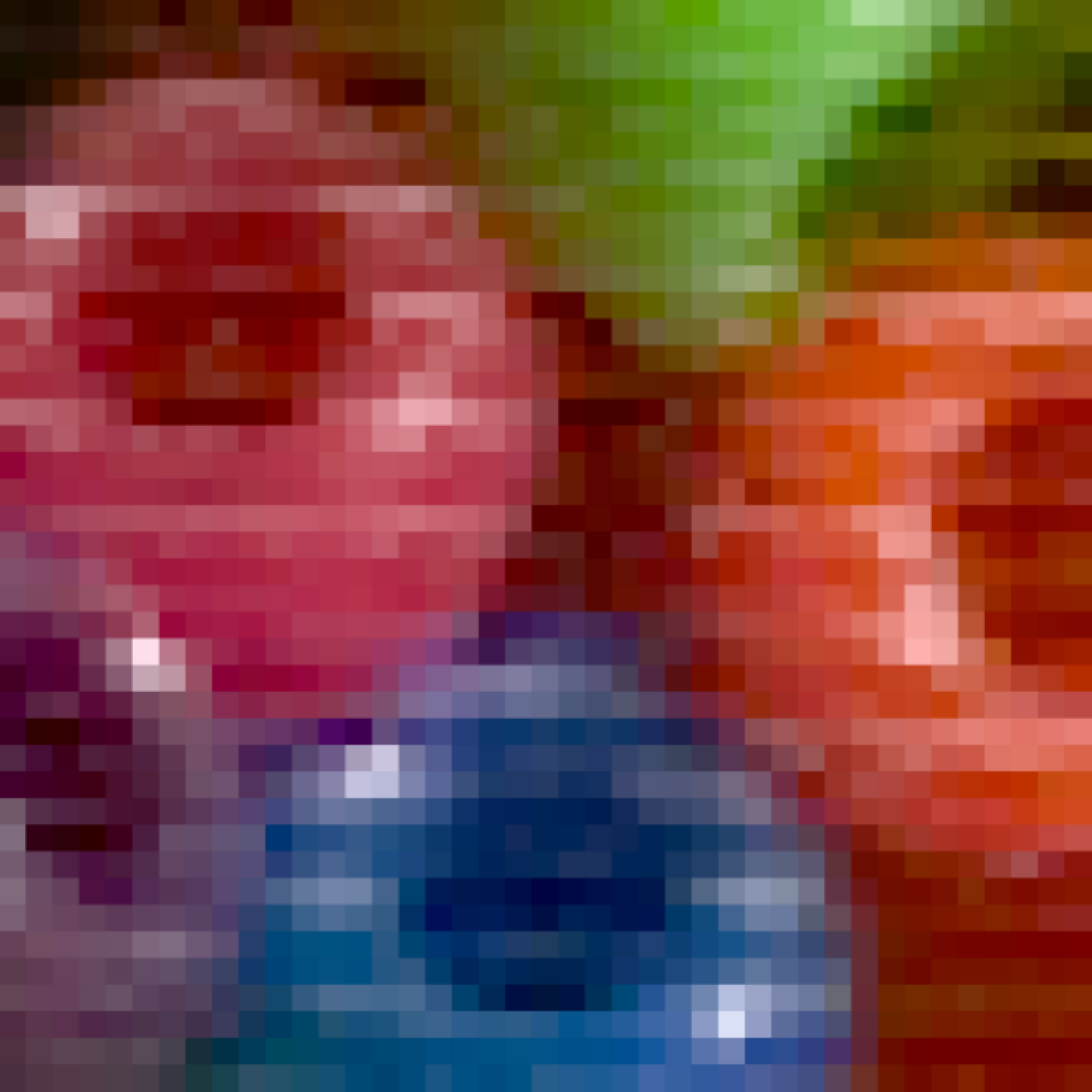}%
  \label{fig:beads_PPID}}
  \hfil
  \subfloat[Proposed]{\includegraphics[width=0.2\linewidth]{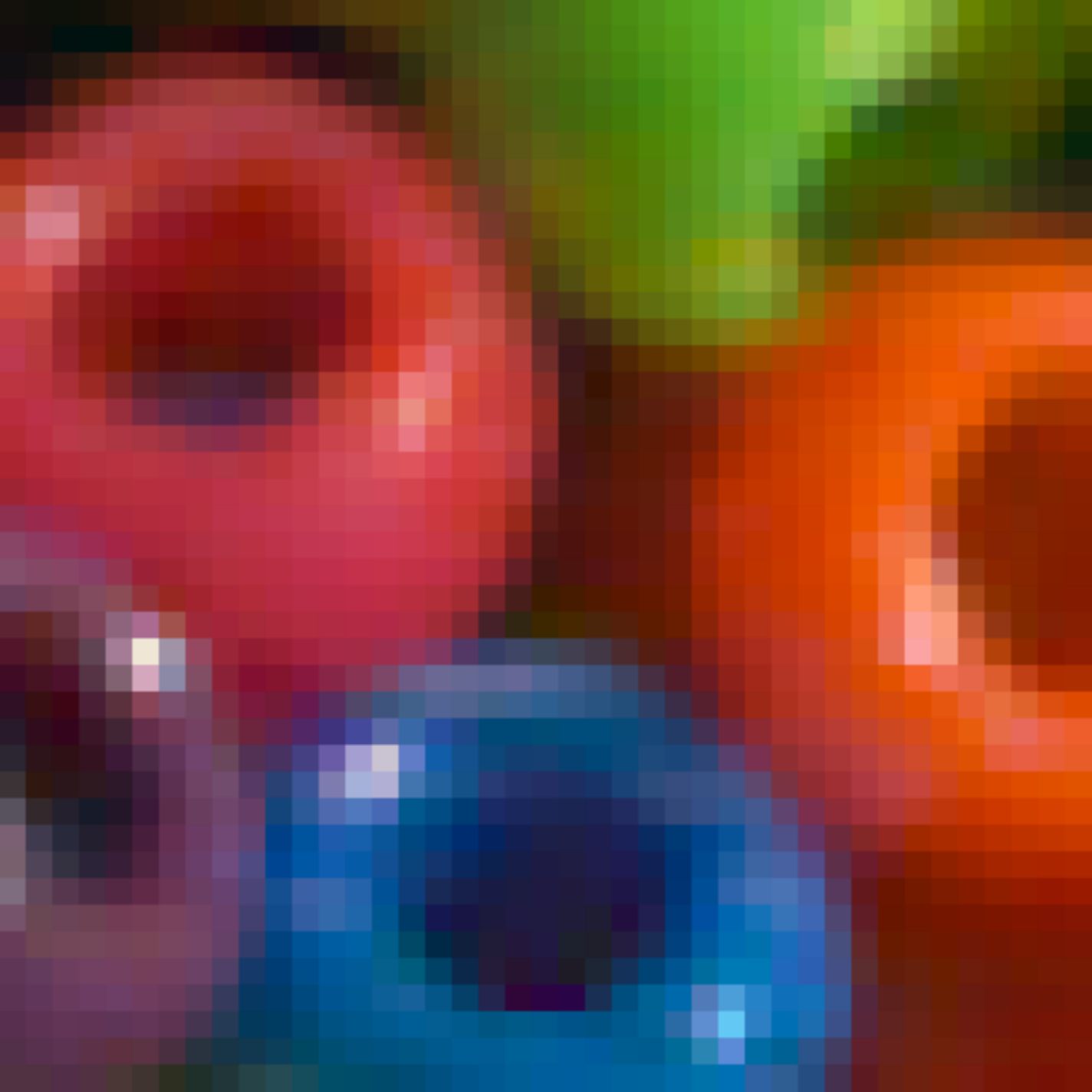}%
  \label{fig:beads_proposed}}
  }  
  \caption{Demosaicked images of ``beads.''}
 \label{fig:beads_demosaicked}
\end{figure*}

\begin{figure*}[tb]
 \centerline{
  \subfloat[Original]{\includegraphics[width=0.2\linewidth]{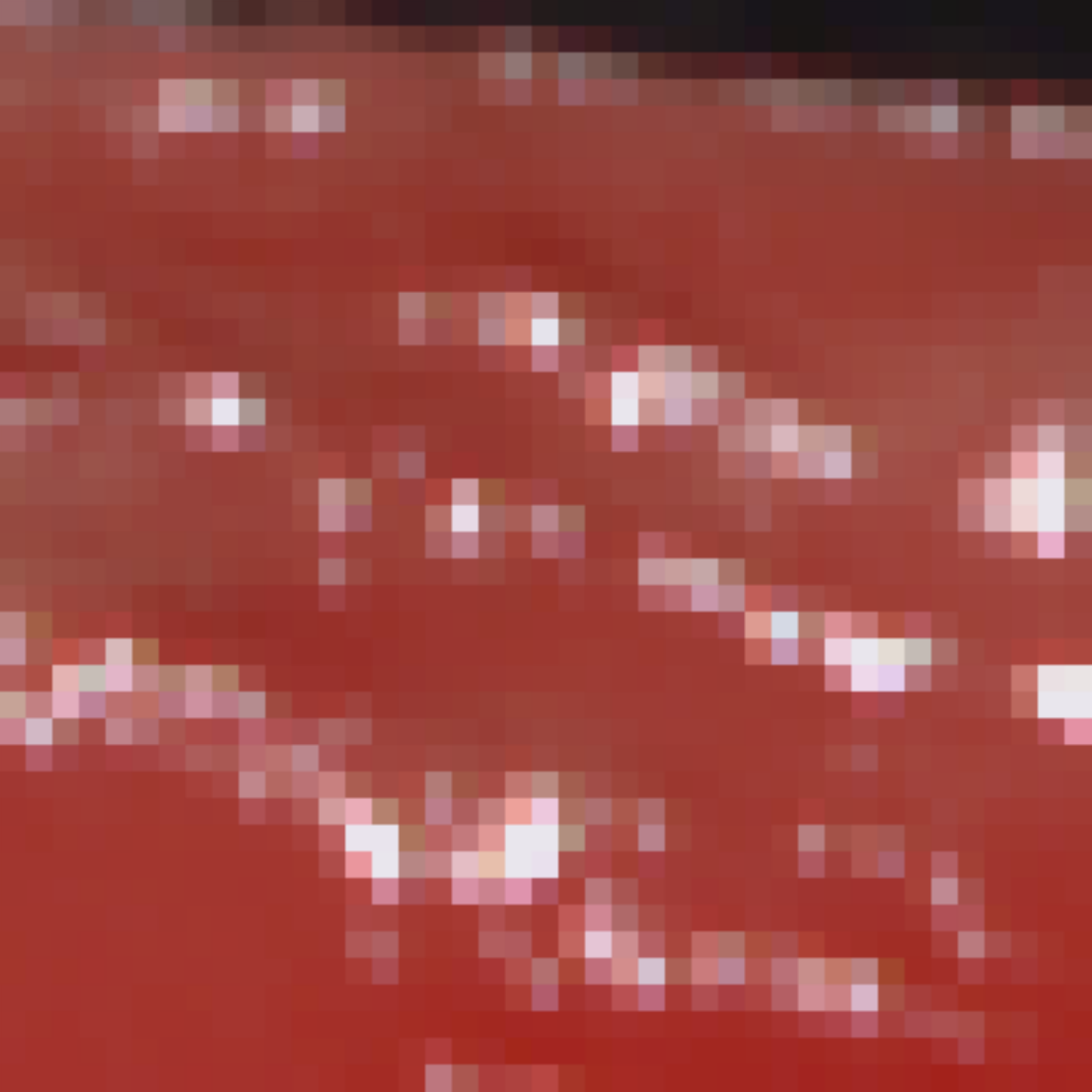}%
  \label{fig:fake_and_real_tomatoes_original}}
  \hfil
  \subfloat[Bilinear]{\includegraphics[width=0.2\linewidth]{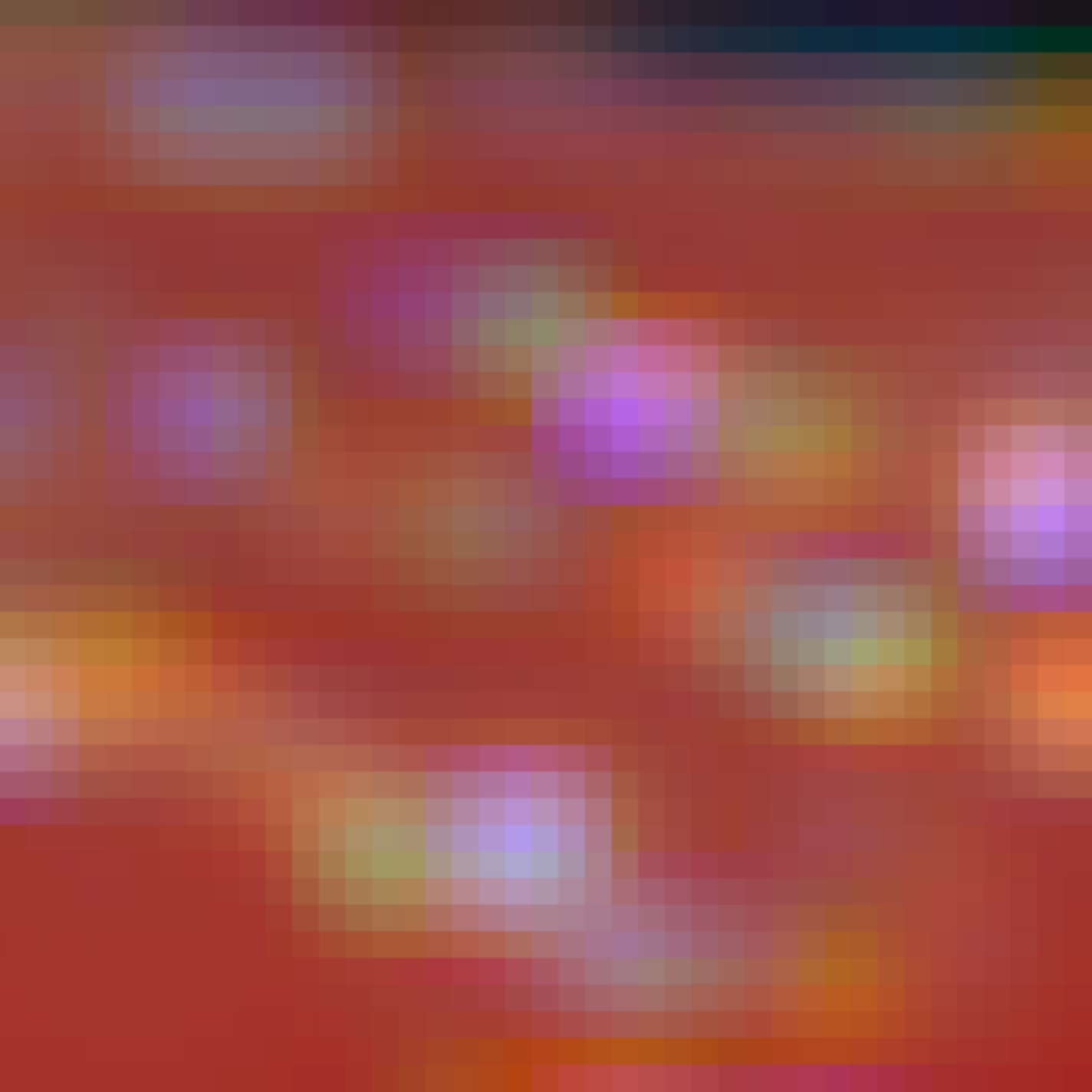}%
  \label{fig:fake_and_real_tomatoes_bilinear}}
  \hfil
  \subfloat[PPID]{\includegraphics[width=0.2\linewidth]{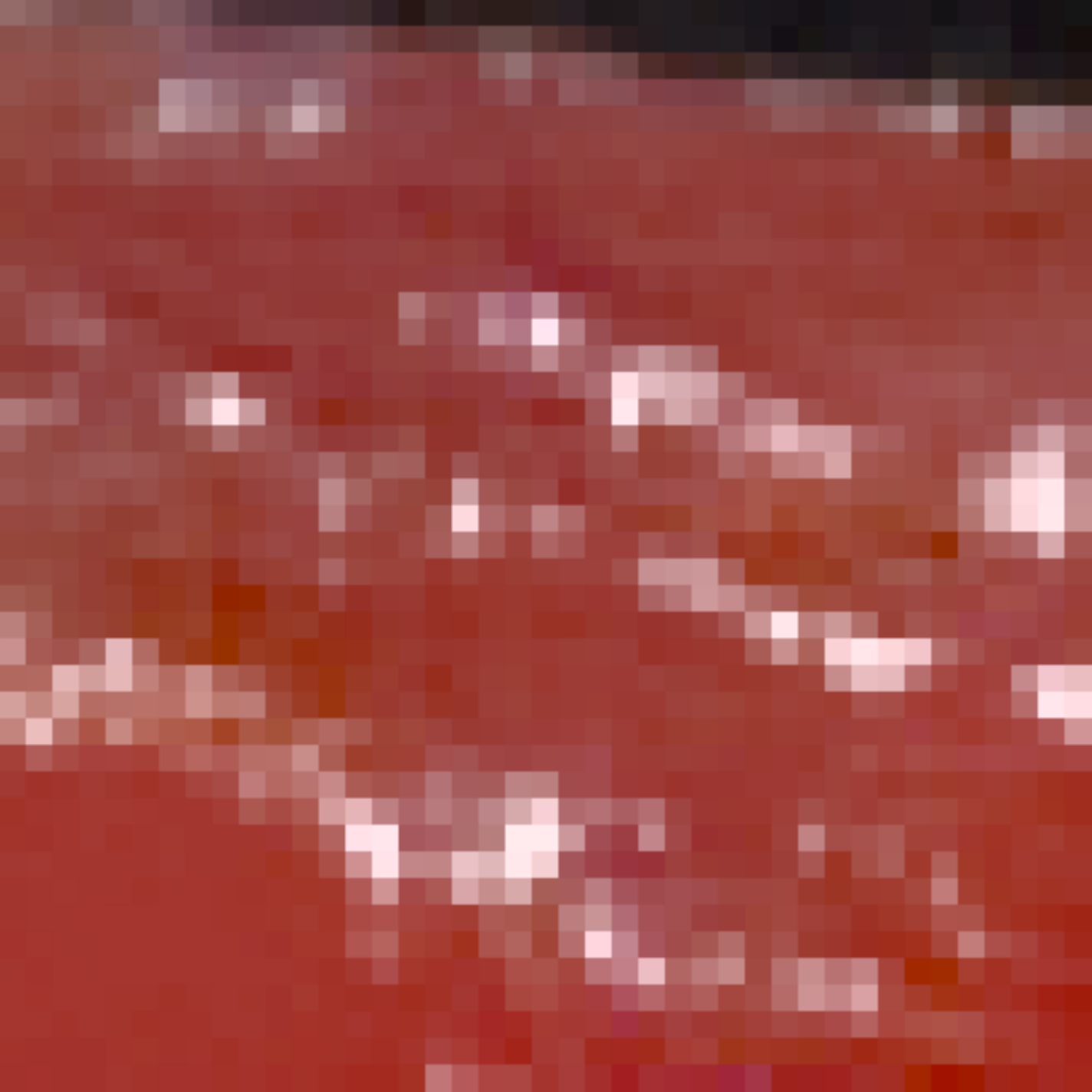}%
  \label{fig:fake_and_real_tomatoes_PPID}}
  \hfil
  \subfloat[Proposed]{\includegraphics[width=0.2\linewidth]{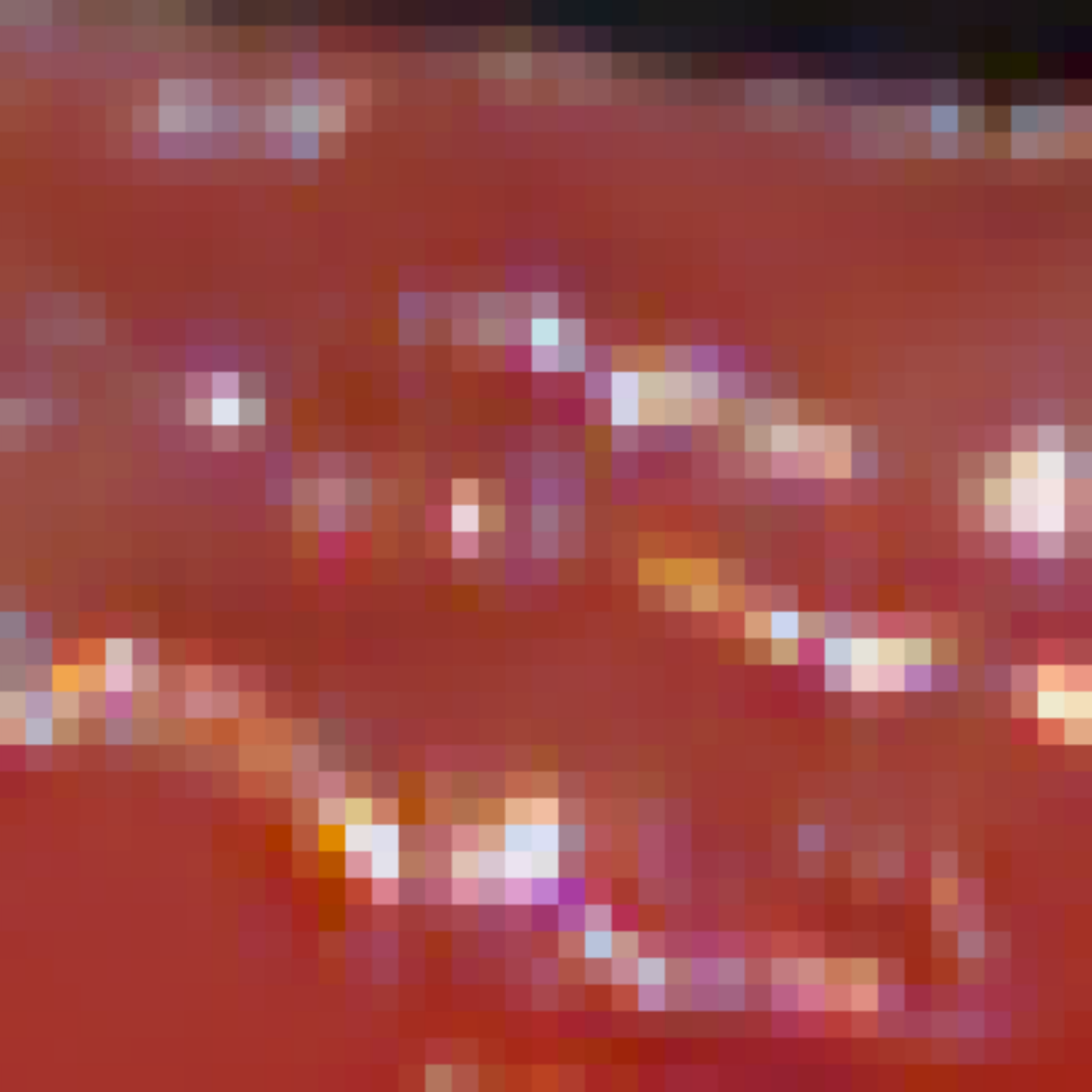}%
  \label{fig:fake_and_real_tomatoes_proposed}}
  }  
  \caption{Demosaicked images of ``fake and real tomatoes.''}
 \label{fig:fake_and_real_tomatoes_demosaicked}
\end{figure*}

The original and demosaicked images are presented in \figsssref{fig:fake_and_real_peppers_demosaicked}{fig:clay_demosaicked}{fig:beads_demosaicked}{fig:fake_and_real_tomatoes_demosaicked}. In \figref{fig:fake_and_real_peppers_demosaicked}, the letters are strongly blurred in the bilinear interpolation results, but readability is significantly improved by the proposed method.
Therefore, the blur in the results of the first step of the proposed method was improved in the second step. In \figref{fig:clay_demosaicked}, an artifact occurred along the edge in the PPID results, whereas the demosaicked image from the proposed method contains no such artifact. This artifact can be seen also in \figref{fig:beads_demosaicked}(c). Therefore, we can conclude that the proposed method suppressed the artifact along the edge. In \figref{fig:fake_and_real_tomatoes_demosaicked}, a color artifact can be seen in the demosaicked image from the proposed method, but cannot be seen in the image from PPID. Because a strong color artifact occurred in the image resulting from bilinear interpolation, the second step of the proposed method was unable to completely remove the artifact. Fixing this issue will require improving our proposed network in the future.

The second step of the proposed method utilizes shortcut connections as a result of residual estimation. If the shortcut connections in modules one to five are removed, the average PSNR of the proposed method decreases by 1.28 dB. Furthermore, in the case where all shortcut connections are removed, this degradation increases to 5.05 dB. Therefore, the ResNet structure performs effectively in the proposed network. Additionally, we can conclude that the 3D-CNN provides an advantage in terms of feature representation of a multispectral data cube because the PSNR of the proposed method decreased by 1.19 dB when the convolution process was changed to a 2D process. Based on these results, the proposed method based on ResNets and a 3D-CNN is suitable for multispectral demosaicking.

%%%%%%%%%%%%%%%%%%%%%%%%%%%%%%%%%%%%%%%%%%%%%%%%%%%%%%%%%%%%%%%%%%%%%%%%%%%%%%%%
\section{Conclusions} \label{sec:level4}
We proposed a novel demosaicking for MSIs utilizing a deep CNN with ResNets and 3D convolution. The proposed method achieved higher PSNRs than conventional multispectral demosaicking methods and can suppress artifacts. Because some demosaicked images from the proposed method still contained color artifacts, we must improve the network structure (e.g., change from bilinear interpolation to PPID in the first step or increase the number of layers) in the future.

%%%%%%%%%%%%%%%%%%%%%%%%%%%%%%%%%%%%%%%%%%%%%%%%%%%%%%%%%%%%%%%%%%%%%%%%%%%%%%%%
\section*{Acknowledgement}
This work was supported by JSPS KAKENHI Grant Number 18K11368.

%%%%%%%%%%%%%%%%%%%%%%%%%%%%%%%%%%%%%%%%%%%%%%%%%%%%%%%%%%%%%%%%%%%%%%%%%%%%%%%%

\end{document}